\documentclass[aps,pra,longbibliography,showkeys,showpacs,groupedaddress,twocolumn,10pt]{revtex4-1}
\usepackage{amssymb}
\usepackage{graphicx}
\usepackage{dcolumn}
\usepackage{bm}
\usepackage{amsmath}
\usepackage{subfigure}
\usepackage{float}
\usepackage{color}
\usepackage{float}
\usepackage{hhline}

\begin{document}
\title{Cs $62D_J$ Rydberg-atom macrodimers formed by long-range multipole interaction}
\author{Xiaoxuan Han$^{1,2}$}
\author{Suying Bai$^{1,2}$}
\author{Yuechun Jiao$^{1,2}$}
\author{Liping Hao$^{1,2}$}
\author{Yongmei Xue$^{1,2}$}
\author{Jianming Zhao$^{1,2}$}
\thanks{Corresponding author: zhaojm@sxu.edu.cn}
\author{Suotang Jia$^{1,2}$}
\author{Georg Raithel$^{1,3}$}
\affiliation{$^{1}$State Key Laboratory of Quantum Optics and Quantum Optics Devices, Institute of Laser Spectroscopy, Shanxi University, Taiyuan 030006, China}
\affiliation{$^{2}$Collaborative Innovation Center of Extreme Optics, Shanxi University, Taiyuan 030006, China}
\affiliation{$^{3}$ Department of Physics, University of Michigan, Ann Arbor, Michigan 48109-1120, USA}
\date{\today}

\begin{abstract}
Long-range macrodimers formed by $D$-state cesium Rydberg atoms are studied in experiments and calculations.
Cesium $[62D_{J}]_2$ Rydberg-atom macrodimers, bonded via long-range multipole interaction, are prepared by two-color photo-association in a cesium atom trap. The first color (pulse A) resonantly excites seed Rydberg atoms, while the second (pulse B, detuned by the molecular binding energy) resonantly excites the Rydberg-atom macrodimers below the $[62D_{J}]_2$ asymptotes. The molecules are measured by extraction of auto-ionization products and Rydberg-atom electric-field ionization, and ion detection. Molecular spectra are compared with calculations of adiabatic molecular potentials. From the dependence of the molecular signal on the detection delay time, the lifetime of the molecules is estimated to be about 6~$\mu$s.
\end{abstract}
\pacs{32.80.Ee, 33.20.-t, 34.20.Cf}
\maketitle

Recently molecules involving one or more Rydberg excitations have attracted considerable attention due to their unusual properties, which include exotic adiabatic potentials, vibrational levels that reveal details of the binding potentials, and permanent dipole moments. Two kinds of Rydberg molecules with distinct binding mechanisms have been observed, Rydberg-ground molecules and Rydberg-Rydberg macrodimers.
Rydberg-ground molecules,
consisting of a Rydberg atom bound to a ground-state atom via a
low-energy electron scattering mechanism, have been theoretically predicted~\cite{Greene,Lesanovsky} and experimentally observed for Rydberg $S$-states~\cite{V. Bendkowsky,Bendkowsky2010}, $P$-states~\cite{M. A. Bellos}, and $D$-states~\cite{D. A. Anderson,A. T. Krupp}, and for high-angular-momentum states~\cite{J. Tallant, Niederprum2016, Kleinbach2017,D. Booth,T. Niederprum}.
Molecules consisting of two Rydberg atoms, bound by long-range electrostatic interactions, have been predicted~\cite{Boisseau} and observed in cold atomic gases~\cite{Overstreet,Deiglmayr,H}. These Rydberg-Rydberg molecules are macrodimers, as their bond lengths are $\sim~4 n^{2}$ and can easily exceed 1~$\mu$m. Since the bond length generally exceeds the LeRoy radius, the exchange interaction is negligible.
Deiglmayr {\sl{et al.}} have prepared Cs macrodimer molecules near the $nS$-$n'F$ and $[nP]_2$ asymptotes for 22$\leq n \leq 32$~\cite{Deiglmayr}, and $43P$-$44S$~\cite{H} molecules bound by long-range dipolar interaction.

\begin{figure}[ht]
\vspace{-1ex}
\centering
\includegraphics[width=0.5\textwidth]{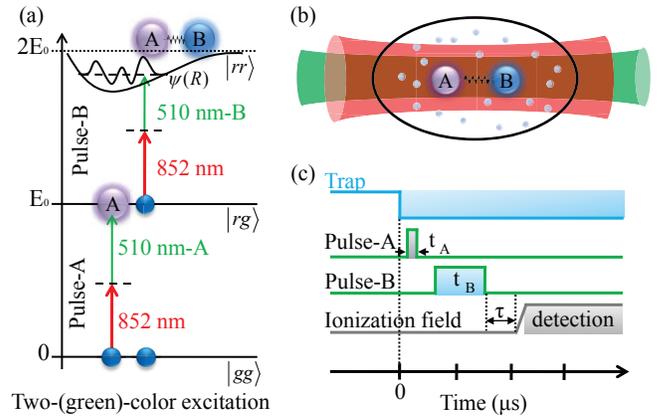}
\vspace{-6ex}
\caption{Level diagram and sketch of a vibrational wave-function (a) and schematic of the experiment (b) for two-color double-resonant excitation of $[62D_{J}]_2$ Rydberg-atom macrodimers.
The 852~nm and 510~nm Rydberg excitation beams counterpropagate through a cold Cs atom cloud. The pulse pair A resonantly excites seed Rydberg atoms (atom A).
To study Rydberg-atom macrodimers, the frequency of the 510-nm component of the pulse pair B is scanned over a range of $\pm$70~MHz relative to the atomic resonance. (c) Timing sequence. After switching off the MOT beams, we sequentially apply the A- and B-pulses.  An optional wait time $\tau \lesssim 40~\mu$s between the B-pulses and detection allows us to study the decay of Rydberg atoms and molecules.}
\end{figure}

Here we observe Cs $[62D_{J}]_2$ macrodimers on the red-detuned side of the 62$D_{J}$ atomic resonances. We excite the molecules via two-step, two-color photoassociation using two sets of laser pulses, as sketched in Fig.~1(a).
A small number of seed Rydberg atoms, denoted as A-atoms, is resonantly excited from the ground state using a laser-pulse pair labeled A. The A-pulses are two-photon resonant with the interaction-free Rydberg level $|r\rangle = $ 62$D_{J}$.
The 510-nm component of the second laser pulse pair, labeled B, is detuned
relative to that of pulse pair A by an amount equal to the molecular binding energy.
The 852-nm components of pulse pairs A and B have the same frequency.
The B-pulses excite Rydberg atoms close to the seed atoms (A-atoms), at a distance where metastable $[62D_{J}]_2$ macrodimers exist. Owing to the doubly-resonant character of this two-color photoassociation scheme, the excitation rate is greatly enhanced in comparison with that of single-color photoassociation.

The experiment is performed in a Cs magneto-optical trap (MOT) with a
temperature $\sim$~100~$\mu$K. After switching off the MOT beams, we apply the photoassociation pulses A and B, as sketched in Fig.~1(a) and (b). The MOT magnetic field is always on. The timing diagram is shown in Fig.~1(c).
The lower-transition laser (852~nm, Toptica DLpro, $\thicksim 100$~kHz linewidth) is stabilized to the $|6S_{1/2}, $F$ = 4\rangle$ ($|g\rangle$) $\to$ $|6P_{3/2},  $F'$ = 5 \rangle$ ($|e\rangle$) transition using polarization spectroscopy~\cite{Pearman}, and is shifted off-resonance from $|e\rangle$ by 220~MHz using a double-pass acousto-optic modulator (AOM).
The upper-transition laser (510~nm, Toptica TA SHG110, 1~MHz linewidth) is stabilized to 62$D_J$ Rydberg transition using a Rydberg EIT reference signal from a Cs vapor cell~\cite{Jiao}, and double-passed through another AOM. For the pulse pair A, the 510-nm AOM frequency is set to resonantly excite the seed Rydberg atoms (A-atoms). During the subsequent B pulse, the 510-nm AOM frequency is detuned and scanned to resonantly photoassociate the B- to the A-atoms, to form molecules.
During the scan, the B-pulse laser power is held fixed using a PID~\cite{Jiao} feedback loop that controls the RF power supplied to the 510-nm AOM. The 852-nm laser has a power of $\thicksim$220~$\mu$W and Gaussian waist of $\omega_{852}$ $\simeq$ 80~$\mu$m. The
510-nm beam has a waist $\omega_{510}$ $\simeq$ 40~$\mu$m at the MOT center.
The excitation region is surrounded by three pairs of field-compensation electrodes, which allow us to reduce stray electric fields to less than 250~mV/cm, via Stark spectroscopy of Rydberg levels. Rydberg atoms and molecules are detected using the electric-field ionization method (ionization ramp rise time 3~$\mu$s). Alternatively, ions formed by Penning ionization of Rydberg atoms or molecules can be extracted with a smaller electric field. The extracted ions are detected with a microchannel plate (MCP) detector. Spontaneously formed ions and field-ionized
Rydberg atoms are discriminated by their time of flight to the MCP.
An ion lens, consisting of three potential grids along the ion path to the MCP, is used to collimate the ions onto the MCP and to thereby optimize the ion collection efficiency.

In our theoretical model of Rydberg-atom macrodimers, we consider the multipole interaction, $V_{int}$, in the product space of two Rydberg atoms labeled $A$ and $B$~\cite{Schwettmann,Deiglmayr},

\begin{eqnarray}
\widehat{V}_{int}&=& \sum_{q=2}^{q_{max}} \frac{1}{R^{q+1}} \sum_{\substack{L_{A}=1 \\ L_B=q-L_A}}^{q_{max}-1}
\sum_{\Omega=-L_{<}}^{L_{<}}f_{AB\Omega}
\hat{Q}_{A} \hat{Q}_{B} \\
f_{AB\Omega}&=&\frac{(-1)^{L_{B}} \, (L_{A}+L_{B})!}{\sqrt{(L_{A}+\Omega)!(L_{A}-\Omega)!(L_{B}+\Omega)!(L_{B}-\Omega)!}} \nonumber
\end{eqnarray}

where $L_{A}$ and $L_{B}$ are the multipole orders of atoms $\emph{A}$ and $\emph{B}$, and $L_{<}$ is the lesser of $L_{A}$ and $L_{B}$; both $L_{A}, \, L_{B} \ge 1$ because the atoms have no monopole moment. For atom $A$, $\hat{Q}_A=\sqrt{4\pi/(2L_{A}+1)}\hat{r}_{A}^{L_{A}}Y_{L_{A}}^{\Omega}(\hat{\bf{r}}_{A})$, with radial matrix elements and spherical harmonics that depend on the Rydberg-electron position operator $\hat{\bf{r}}_A$. For atom $B$ an equivalent expression applies, with $\Omega$ replaced by $-\Omega$.  The outer sum ends at a maximum order $q_{max}$. We have tested cases up to $q_{max}=6$ and found that $q_{max}=4$ is sufficient to model potentials and level crossings with $\lesssim 1$~MHz accuracy.
For $q_{max}=4$, the terms included are $dd$, $do$, $od$, $dh$, $qq$, and $hd$ interactions, where the first (second) letter stands for atom $A$($B$), and $d$ stands for dipole, $q$ for quadrupole, $o$ for octupole, and $h$ for hexadecupole interaction. We diagonalize the Hamiltonian on a grid of the internuclear separation, $R$. Due to azimuthal symmetry, the sum of the projections of the electronic angular momenta onto the internuclear axis, $M=m_{jA} + m_{jB}$, is conserved.

\begin{figure}[tbp]
\vspace{1ex}
\centering
\includegraphics[width=0.45\textwidth]{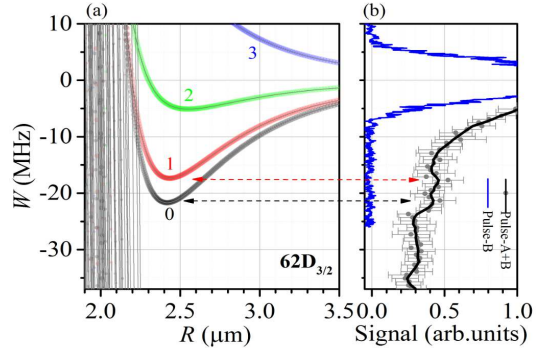}
\vspace{-1ex}
\caption{(a) Calculations of adiabatic potentials (gray lines) for cesium Rydberg  macrodimers $[62D_{3/2}]_2$, for the indicated values of $M$. Excitation rates are proportional to the areas of the overlaid colored circles.
(b) Spectra measured with single-color excitation (pulse-B only; blue line) and $[62D_{3/2}]_2$ macrodimer spectra for two-color excitation (pulses A and B; gray symbols and error bars show original data, black line shows smoothed average). The peak of the single-color spectrum marks the $[62D_{3/2}]_2$ asymptote (detuning $W=0$). Horizontal dashed lines indicate the minima of two $[62D_{3/2}]_2$ molecular potentials.
}
\end{figure}

Figures~2(a) and~3(a) show calculated adiabatic potentials below the $[62D_J]_2$ asymptotes
for $J=3/2$ and 5/2, respectively, for $q_{max}=6$, and for the relevant values of $M$, $\vert M\vert =0, 1, ... , 2 J$. Symbol areas are proportional to oscillator strength for the laser excitation; symbol colors correspond to different $\vert M \vert$-values.
While most adiabatic potentials do not exhibit minima that could support bound Rydberg dimers (gray lines), all values of $M$, except $\vert M\vert=2J$, support one binding potential. The binding potentials carry large oscillator strengths and exhibit minima that are up to $\sim 22$~MHz and $\sim 42$~MHz deep, for $J=3/2$ and 5/2, at distances of $\sim 2.4~\mu$m. Figure~3 also shows several potentials without notable oscillator strength that intersect with the binding potentials. Close inspection shows that most intersections are unresolved (non-adiabatic coupling $\ll 1$~MHz), except for $\approx 2$-MHz wide anti-crossings at -27~MHz in both the $M=0$ and $M=1$ binding potentials, a $\approx 1$-MHz wide one at $-39$~MHz in $M=1$, and a sub-MHz wide one at $-35.5$~MHz in $M=2$ (all for $J=5/2$).

The black lines in
Figs.~2(b) and 3(b) show two-color photoassociation spectra recorded below the $[62D_J]_2$ asymptotes. The 852-nm beam power is 210~$\mu$W, and the 510-nm powers are 5.7~mW for $J=3/2$ and 1.2~mW for $J=5/2$. The duration of the A-pulses is 0.5~$\mu$s for $J=3/2$ and $0.2~\mu$s for $J=5/2$, and that of the B-pulses 6.0~$\mu$s (for both $J$). The 510-nm powers and A-pulse durations are chosen differently to (partially) compensate the different optical excitation matrix elements for the $62D_J$ Rydberg levels. Rydberg atoms and molecules are field-ionized and detected immediately after the B-pulses. At detunings $\vert W \vert \gtrsim 10~$MHz the $B$-pulses only generate signal when the much shorter A-pulses are on, {\sl{i. e.}} if the sample is seeded with A-Rydberg-atoms. This demonstrates that the doubly-resonant photoassociation procedure is very efficient in generating Rydberg-atom pairs at large detunings.

\begin{figure}[t]
\vspace{1ex}
\centering
\includegraphics[width=0.45\textwidth]{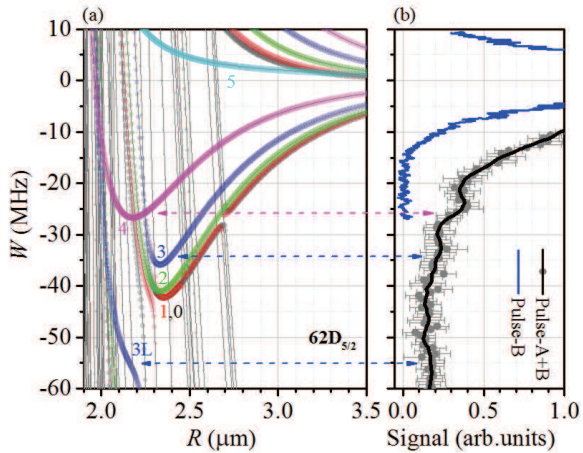}
\vspace{-1ex}
\caption{(a) Calculations of adiabatic potentials for $[62D_{5/2}]_2$ macrodimers, and (b) measured spectra. The figure is analogous to Figure~2.
}
\end{figure}

The measured photoassociation signals in Figs.~2(b) and 3(b) include a structure-less part that rapidly drops off as a function of detuning; this part is attributed to van-der-Waals-type interaction at distances and conditions away from the aforementioned adiabatic-potential minima. In both $J$-cases, the measured spectra exhibit additional peaks near the minima of several of the highlighted adiabatic potentials. Most notable are the signals attributed to the $M=3, 4$ potentials of $J=5/2$ and the $M=0,1$ potentials of $J=3/2$. These potentials also have the largest numbers of bound vibrational states, as shown next. We believe that the other potentials fail to produce significant signals because they support a lesser number of bound vibrational states ($M=0,1,2$ for $J=5/2$), or because their binding energy is so small that the signals are obscured by the atomic background signal ($M=2$ for $J=3/2$). The $M=3$-case for $J=5/2$ exhibits an adiabatic potential that has a ``knee'' near $-55~$MHz. The knee has large oscillator strength and may produce some signal attributed to dissociating atom pairs (see $3L$ label in Fig.~3).

To gauge the importance of a binding potential, $V(R)$, in producing a signal indicative of metastable Rydberg dimers, the number of vibrational states below an energy $V_{top}$ can be estimated semiclassically as $N(V_{top}) = 2 \int_{R_i}^{R_o} p(R) dR / h + 1/2$, with momentum $p=\sqrt{2 \mu (V_{top}-V(R))}$, inner and outer classical turning points $R_i(V_{top})$ and $R_o(V_{top})$, and effective mass $\mu = M_{Cs}/2 \approx 66.5~u$. For the results given in Table I, we set $V_{top}$ equal to the energies of the aforementioned anti-crossings in the adiabatic potentials. Vibrational states above these anti-crossings are likely unstable; we assume that they decay into fast ions that avoid detection after the fairly long B-pulse. In the cases where there are no relevant anti-crossings we use $V_{top}=V_{min}/2$. The binding adiabatic potentials with the largest numbers of metastable vibrational states and DOS are expected to produce the strongest spectroscopic signals. The results in Figs.~2 and~3 and Table I support this interpretation.

\begin{table}
    \caption{The rows show the statistical weight vs $\vert M \vert$, the measured and calculated minima of the binding adiabatic potentials of Cs $[62D_{J}]_2$, $V_{min}$, the energy at which the number of
    vibrational levels is evaluated, $V_{top}$, the number of energy levels, $N$, and the average density of states (DOS) between
    $V_{min}$ and $V_{top}$. Energies are in MHz, and DOS in MHz$^{-1}$.}
    \begin{center}
\begin{tabular}{|r|r|r|r|r|r|}
  \hline
   $M$ = & 0 & 1 & 2 & 3 & 4 \\
  \hline
   stat. weight & 1 & 2 & 2 & 2 & 2 \\
  \hhline{|=|=|=|=|=|=|}
   $[62D_{3/2}]_2$ & ~ & ~ & ~ & ~ & ~ \\
  \hline
$V_{min}$, Exp.  & -22 $\pm$ 2 & -17 $\pm$ 2 & - & - & - \\
$V_{min}$, Theo. & -21.7 & -17.4 & -5.1  & -    & -  \\
  \hline
$V_{top}$                & $V_{min}/2$     & $V_{min}/2$     & $V_{min}/2$     &   -  &  -   \\
  \hline
  $N$                  & 61    &   55  & 31    &   -  &   -  \\
  \hline
  DOS                & 5.6   &   6.3 & 12.2   &   - &  -   \\
  \hhline{|=|=|=|=|=|=|}
  $[62D_{5/2}]_2$ & ~ & ~ & ~ & ~ & ~ \\
  \hline
$V_{min}$, Exp.   & -40 $\pm$ 3 & -40 $\pm$ 3 & -40 $\pm$ 3 & -32 $\pm$ 2 & -25 $\pm 2$  \\
$V_{min}$, Theo.  & -42.2 & -42.3 & -41.0 & -35.8 & -26.7  \\
  \hline
$V_{top}$               & -28.1 & -39.6 & -35.6 &   $V_{min}/2$   &  $V_{min}/2$     \\
  \hline
  $N$                  & 44    &    7  & 15    &   58  &   61    \\
  \hline
  DOS                & 3.1   &   2.6 & 2.8   &   3.2 &  4.6   \\
  \hline
\end{tabular}
    \end{center}
\end{table}

Our theoretical model also yields the two-body excitation rate of two-color photoassociation of the binary molecules (methods similar to~\cite{Schwettmann,Deiglmayr}). Since our sample is not optically pumped nor are the atoms spatially pre-arranged, the calculated rates are averaged over a random spatial distribution of positions and angles between internuclear axis and laboratory frame (defined by laser polarizations), and summed over all $M$. Since $M$-values of same magnitude and opposite sign have the same spectra, the non-zero $\vert M \vert$-values carry a weight of 2 (see Table I). Figure~4 shows comparisons between
measured and calculated two-color excitation spectra on the red-detuned sides of $[62D_J]_2$. The structures in the calculated and measured spectra, indicative of the production of macrodimer Rydberg molecules, are in good qualitative agreement for both cases of $J$.

\begin{figure}[htbp]
\vspace{-1ex}
\centering
\includegraphics[width=0.4\textwidth]{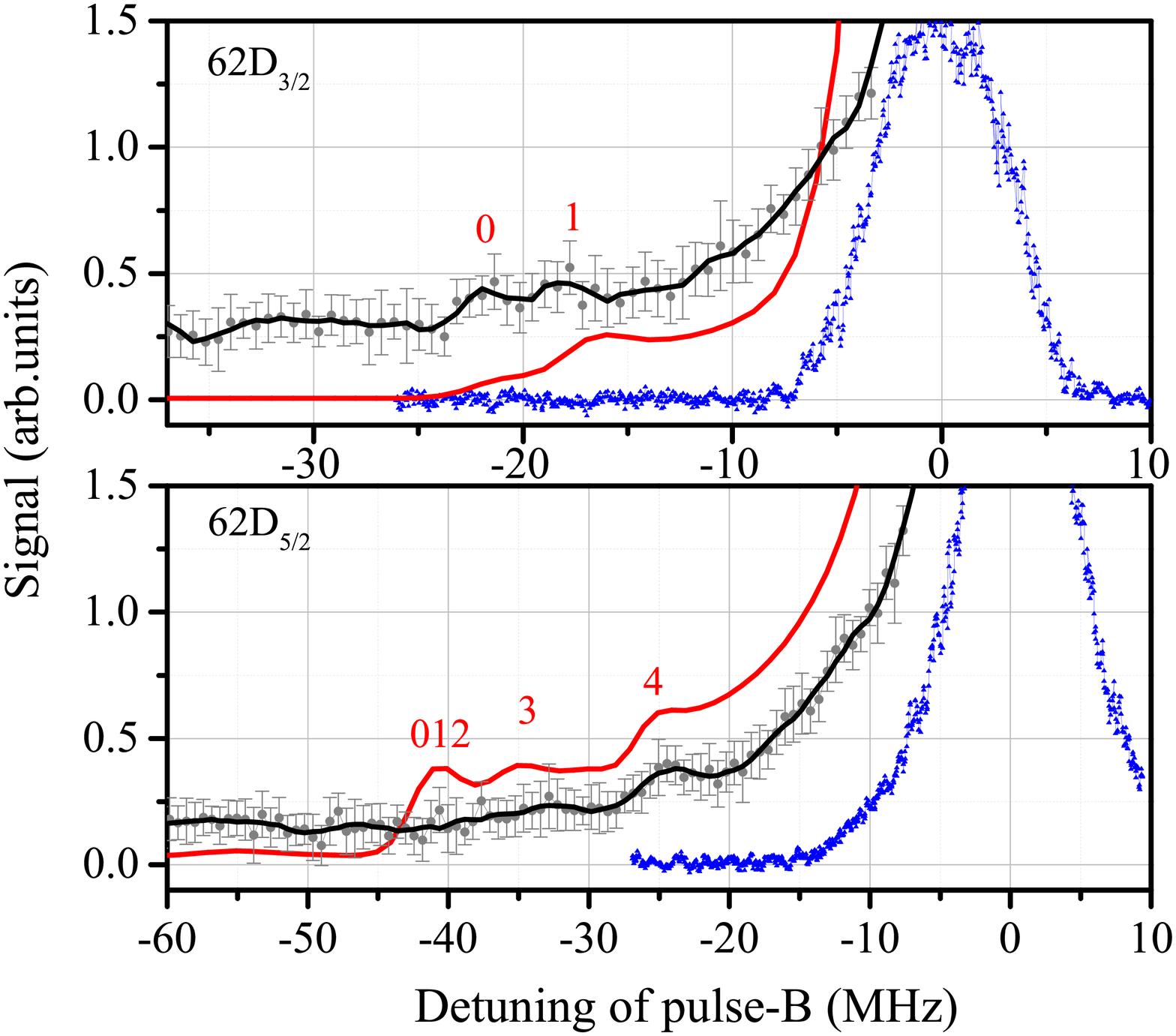}
\vspace{-1ex}
\caption{ Measurements and calculations (red lines) of doubly-resonant two-color photoassociation spectra for atom density $10^8$~cm$^{-3}$ and laser-excitation FWHM of 3~MHz  for $[62D_{J}]_2$ Rydberg molecules vs pulse-B detuning.
The data points (gray symbols with error bars) are measurements with seed pulse A on, the black lines show smoothed averages. The narrow signals (blue lines) are atomic reference signals with the seed pulse off. Numbers indicate the $M$-values and the energy positions of the binding adiabatic potentials.}
\end{figure}

\begin{figure}[htbp]
\vspace{1ex}
\centering
\includegraphics[width=0.4\textwidth]{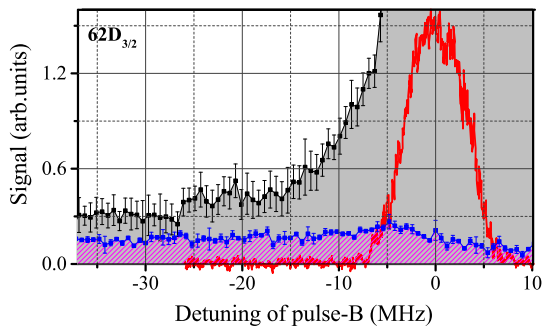}
\vspace{-1ex}
\caption{Spectra of $[62D_{3/2}]_2$ macrodimers obtained with field ionization (0-$\mu$s delay; black line) and ion extraction (12-$\mu$s delay; blue line). For comparison, the atomic Rydberg signal is also plotted (pulse-B only; red line).}
\end{figure}

Rydberg macrodimers are likely to undergo auto-ionization (Penning ionization out of a bound Rydberg-Rydberg state)~\cite{Viteau,Merkt} due to the proximity of the constituents. This allows for detection of Rydberg-atom macrodimers via ion extraction, leading to a signal with a greatly reduced single-atom background. In this method, after photoassociation we insert a wait time, $\tau$ =12~$\mu$s, during which the Rydberg molecules may undergo Penning ionization. We then apply an electric field that is less than the ionization field for the selected atomic state; the electric field is, however, sufficient to collect any spontaneously formed ions onto the MCP. The blue curve in Fig.~5 shows the molecular spectrum below the $[62D_{3/2}]_2$ asymptote obtained with the ion extraction method (free-ion spectrum). For comparison, we also display the molecular signal obtained with the field-ionization method (black curve) and the one-color Rydberg excitation signal (red line). At large detunings, the strength of the free-ion molecular signal is about half of the field-ionization molecular signal. Thus, a large fraction of the Rydberg molecules excited by the A- and B-pulse sequence Penning-ionize with high probability during the waiting time. Close to the atomic resonance, the free-ion signal is much smaller than the
field-ionization signal, showing that the latter contains a large background due to un-paired atoms that do not Penning-ionize. The free-ion signal is largely free of the atomic background. Figure~5 affirms the high efficiency of the doubly-resonant photoassociation. It is further noted that the free-ion signal drops on the blue side of the atomic resonance. This is expected due to the suppression of Penning ionization caused by repulsive van-der-Waals forces on the blue side of the resonance.

To investigate the lifetime of $[62D_{5/2}]_2$ Rydberg-atom macrodimers, in two sets of measurements we keep the pulse-B detuning fixed, at $W=-25$ and $-36$~MHz, respectively, and vary the wait time $\tau$ (see Fig.~1(c)). A ramped detection electric field is used to separately measure the ions that result from Penning ionization of molecules, and the molecules that do not ionized. The fractions of non-ionized molecules drop exponentially with respective fitted decay times of 5.5$~\pm$~0.7~$\mu$s and 6.8$~\pm$~0.9~$\mu$s. These times provide an initial indication for the lifetimes of the $[62D_{5/2}]_2$ molecules.

In conclusion, we have studied Rydberg-atom macrodimers
using two-color photoassociation. The frequency difference between the two colors yields the molecular binding energy.
The measured spectra agree reasonably well with calculated molecular potentials and spectra.
In the future, one may study mixed-parity photoassociation schemes,
realized by simultaneous two- and three-photon Rydberg-atom excitation. A detailed understanding of the
lifetimes of the $D$-state Rydberg-atom molecules will require further study; this includes the role of unresolved sub-MHz non-adiabatic couplings with unbound potentials that may limit the stability of the molecular states.
Also, it has been suggested that Rydberg-atom macrodimers can be used to study vacuum
fluctuations~\cite{L. H. Ford,G. Menezes}, to quench ultracold collisions~\cite{Boisseau}, and to measure correlations in quantum gases~\cite{Overstreet,M. Stecker}.

The work was supported by the National Key R$\&$D Program of China (Grant No. 2017YFA0304203), the National Natural Science Foundation of China (Grants Nos. 11274209, 61475090, 61675123 and 61775124), Changjiang Scholars and Innovative Research Team in University of Ministry of Education of China (Grant No. IRT13076), and the State Key Program of National Natural Science of China (Grant No. 11434007). GR acknowledges support by the National Science Foundation (PHY-1506093) and BAIREN plan of Shanxi province.

\end{document}